\documentclass[10pt]{iopart}
\usepackage[english]{babel}
\usepackage{csquotes}
\usepackage{iopams}
\usepackage{parskip}

\usepackage[square,sort&compress,comma,numbers]{natbib}
\usepackage[font=small,labelfont=bf]{caption} 
\usepackage{amsmath}
\usepackage{enumerate}
\usepackage{amssymb}
\usepackage{mathtools}
\usepackage{gensymb}
\usepackage{enumitem}
\usepackage{multirow}
 \usepackage{indentfirst}
\usepackage{amsmath}
\usepackage[T1]{fontenc}
\usepackage[scaled=0.85]{helvet}
\DeclareTextFontCommand{\ba}{\sffamily}
\usepackage{graphicx}
\usepackage{physicsedited}
\usepackage{wrapfig}
\graphicspath{{}}
\usepackage{fancyhdr}
\usepackage{bm}
\usepackage{geometry}
\usepackage{comment}
\usepackage{amsbsy}
\usepackage{float}
\usepackage[colorlinks = true,
            linkcolor = blue,
            urlcolor  = blue,
            citecolor = blue,
            anchorcolor = blue]{hyperref}
\usepackage{array}
\usepackage{url}
\usepackage[noabbrev]{cleveref}
\newcolumntype{M}[1]{>{\centering\arraybackslash}m{#1}}
\allowdisplaybreaks

\begin{document}
\title[Paper]{Magnetic field orientation dependence of continuous-wave optically detected magnetic resonance with nitrogen-vacancy ensembles}
\author{Pralekh Dubey$^1$, Shashank Kumar$^1$, Chinmaya Singh$^1$, Jemish Naliyapara$^1$, Monish A Poojar$^1$, Harikrishnan K B$^2$, Anshul Poonia$^1$, and Phani Peddibhotla$^1$ \footnote{Present address: Department of Physics, IISER Bhopal, India}}
\address{$^1$ Department of Physics,
Indian Institute of Science Education and Research Bhopal, Madhya Pradesh, India}
\address{$^2$ Department of Physics,
Indian Institute of Science Education and Research Pune, Maharashtra, India}
\ead{phani$@$iiserb.ac.in}

\begin{abstract}\\
Continuous-wave optically detected magnetic resonance (CW-ODMR) measurements with nitrogen-vacancy (NV) spins in diamond are used for sensing DC magnetic fields from nearby magnetic targets. However, this technique suffers from ambiguities in the extraction of the magnetic field components when resonances due to different NV orientation classes overlap with each other. Here, we perform detailed experimental and theoretical studies of such effects on NV ensembles experiencing low bias magnetic fields. In particular, through symmetry considerations, we systematically examine the ODMR response of different NV orientation classes as a function of the orientation of the magnetic field vector. 
Our studies are of importance for performing a careful and detailed analysis of the ODMR spectra in order to infer the vector magnetic field information. Our results find application in the studies of magnetic samples that require a low applied bias field and also can be potentially adapted to defect spins in other solid-state systems.

\end{abstract}
\noindent{\it Keywords\/}: nitrogen-vacancy center, magnetometry, diamond, quantum sensing


\setcounter{page}{1}

\section{Introduction}
\label{section:Introduction}
\linespread{1.4}\selectfont
Over the past two decades, the field of magnetometry has experienced significant advancements, primarily driven by the emergence of quantum sensors \cite{Degen:2017}. A prominent quantum sensor in this field is the negatively charged nitrogen-vacancy (NV) center in diamond, a solid-state defect known for its unique properties. The NV center is a point defect, which is formed by a substitutional nitrogen atom adjacent to a vacancy in the diamond lattice. It is a spin-1 system, which can be initialized, readout, and controlled by optical means. This accessibility of the spin state, combined with its long coherence times even at room temperature, and a wide operational range, makes it useful for magnetometry applications. NV centers are distinguished by their remarkable properties, including their ability to measure magnetic fields with nanoscale spatial resolution and high sensitivity \cite{Taylor_2008}. Recent experiments with diamond defects have demonstrated magnetic sensitivity in the order of femtotesla range, \cite{Silani:2023} comparable to the sensitivities achieved with the alkali-metal vapour magnetometers \cite{Bloom:1962, Budker:2007, Savukov:2005} and superconducting quantum interference devices (SQUID) \cite{Greenberg:1998,Storm:2017, Qiu:2007}. Additionally, the NV centers are sensitive to stress \cite{Barson:2017, Kehayias:2019, Hseih:2019}, temperature \cite{Acosta:2010,Abrahams:2023}, and electric fields \cite{Dolde:2011,Block:2021}, making them a leading platform for multifaceted sensing applications. \\ 
\begin{figure}[h]
\centering
\includegraphics[width=0.9\textwidth]{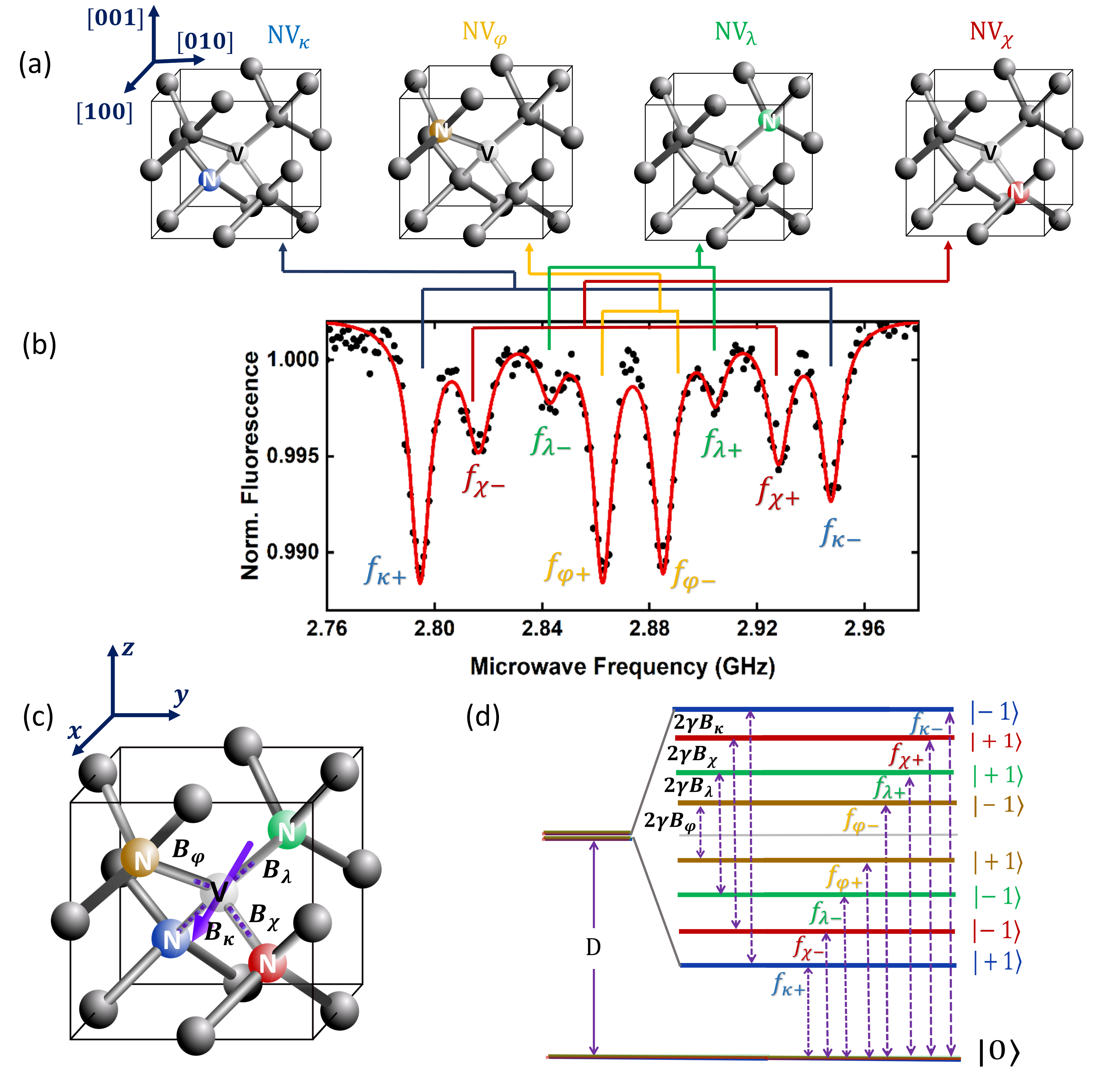}
\caption{
(a) The diamond lattice hosts four possible NV center orientations, labeled as NV$_\kappa$, NV$_\varphi$, NV$_\lambda$, and NV$_\chi$. For the case of dense NV center ensembles in a single-crystal diamond, the NV centers are equally distributed among the four orientations. For small static magnetic fields (<4 mT), the spin quantization axis of the NV center is along one of the four possible \(\langle 111 \rangle\) crystallographic directions. (b) ODMR spectrum corresponding to a magnetic field \(\vb*{B}\) that results in distinct projections on all four NV orientations. As a result, the Zeeman splitting \(\Delta{f_{i}} = |f_{i+} - f_{i-}|\) varies for each NV orientation, leading to the appearance of eight dips in the spectrum. (c) Schematic representation of the diamond lattice illustrating how the magnetic field \(\vb*{B}\) projects onto the four NV orientations. The projections depend on the polar (\(\theta\)) and azimuthal (\(\phi\)) angles of \(\vb*{B}\) relative to the diamond crystal coordinate system (\(x, y, z\)). (d) The energy level splittings corresponding to the magnetic field configuration in Fig. 1 (c). Transition frequencies correspond to the dips visible in the ODMR spectrum.  
}
\label{fig:heatmap}
\end{figure}
\hspace*{10pt} Conventional NV magnetometry employs continuous-wave optically detected magnetic resonance (CW-ODMR) technique for measuring magnetic fields, owing to its operational simplicity. In a dense ensemble of NV centers, the symmetry axis of the NV center can align along one of the four possible crystallographic axes \cite{Doherty:2013} and all four NV orientation classes are equally populated. In low magnetic field experiments, each class of NV centers experiences a Zeeman splitting proportional to the projection of the magnetic field along its axis. This results in a characteristic ODMR spectrum that typically exhibits four distinct pairs of resonance dips (or eight resonance dips). However, under certain geometric alignments of the applied magnetic field $\vb*{B}$ and the NV symmetry axes, two or more of the pairs of resonance dips may overlap and/or the resonance dips within a pair may merge (if the projection along the corresponding NV axis is zero), leading to an ODMR spectrum with fewer dips. These features become even more pronounced in CW-ODMR measurements at low bias fields due to the power broadening effects \cite{Dreau:2011}, complicating the accurate extraction of vectorial magnetic field information. \\
\hspace*{10pt} Previous studies have addressed these challenges by introducing large bias fields to avoid overlapping resonances \cite{Chatzidrosos:2021,Glenn:2017} in conventional ODMR spectral measurements, employing advanced Fourier optical processing techniques to resolve the ODMR contributions from each NV orientation \cite{Backlund:2017}, or using carefully engineered sensor arrangements, such as fiber-integrated magnetometers or cavity-enhanced devices, to improve the collection efficiency and signal-to-noise ratio, thereby partially alleviating the discussed effects \cite{ Homrighausen:2024, Chatzidrosos:2021, Chatzidrosos:2017}. However, these solutions often impose complexity or operational limitations that are not always suitable for low-field sensing scenarios. This motivates us to perform a systematic study of the overlapping ODMR peaks in order to understand the ambiguities that may arise in the extraction of the magnetic field components in low to moderate magnetic fields. \\
\hspace*{10pt} In this work, we present a detailed experimental and theoretical investigation of the dependence of the NV ensemble ODMR spectrum on the geometric relationship between the applied magnetic field $\vb*{B}$ and the four NV orientations. We use symmetry considerations to systematically deduce the conditions under which the ODMR spectrum can exhibit a variable number of resonance dips from two to eight resonances. While monitoring the ODMR spectrum as it evolves with changes in the direction of $\vb*{B}$, our studies enable reliable matching of the ODMR resonance peaks with the corresponding NV center families.
\begin{figure}
\centering
\includegraphics[width=1\textwidth]{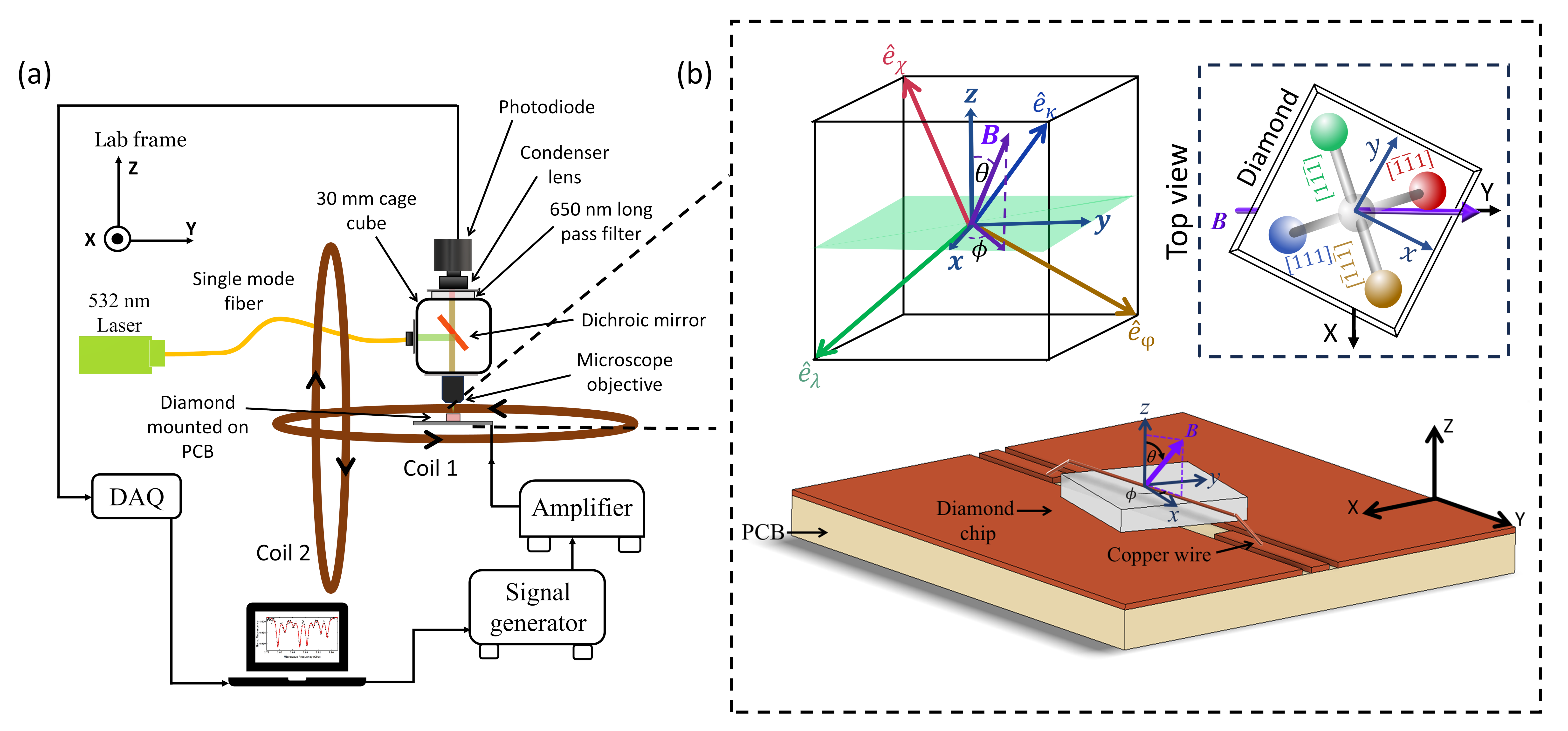}
\includegraphics[width=1
\textwidth]{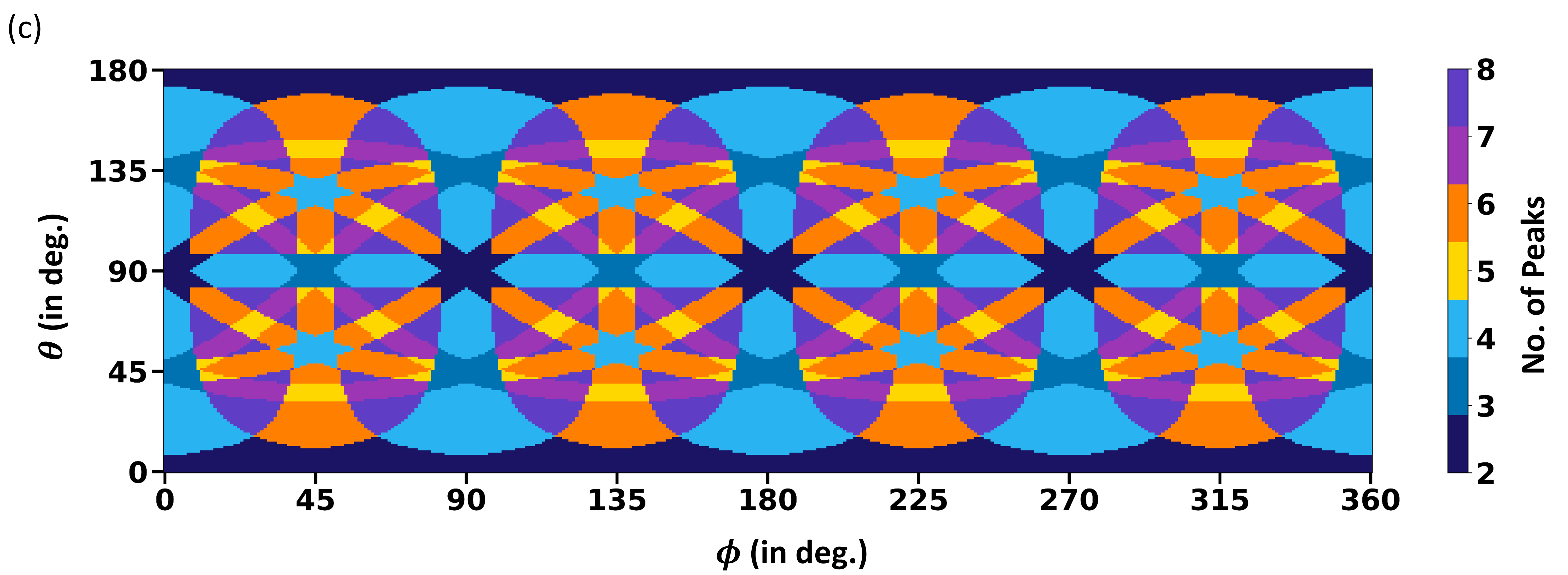}
\caption{
(a) Schematic of the experimental setup, illustrating the optical, electronic, and microwave components. The NV magnetometer setup makes use of a 30 mm cage mount  that integrates all essential optics for fluorescence collection and filtering. A photo-diode converts the NV fluorescence into a voltage signal, which is then processed using a data acquisition (DAQ) system. The electromagnetic coils used for applying a DC magnetic field, labeled "Coil 1" and "Coil 2", are aligned along the Y and Z axes of the laboratory coordinate system, respectively, allowing controlled field application. 
(b) A zoomed-in view of the diamond sample and the PCB which transmits the microwave field. The applied magnetic field \(\vb*{B}\) is depicted with respect to both the laboratory coordinate system (\ba{X, Y, Z}) and the diamond crystal coordinate system $(x,y,z)$, illustrating its orientation in both coordinate systems. The inset provides a top-down view of the diamond crystal, highlighting the projections of the \(\langle111\rangle\) NV axes and their direction relative to the applied magnetic field.
(c) Heat-map depicting the number of observable ODMR dips as a function of the orientation angles \(\theta\) and \(\phi\) for an applied magnetic field of magnitude \(|\vb*{B}| = 3\) mT. Here, any two resonance dips are considered distinct if the separation between them \(\geq 12\) MHz. This is done to account
for the power-broadening effects we experimentally observe in the CW-ODMR spectrum. We note the eight-fold symmetry in the heat-map, which can be attributed to the three-fold mirror symmetry and the three-fold rotational symmetry about the NV axes. 
}
\label{fig:exp_fig}
\end{figure}
\section{Methods and materials}
\label{section:methods}
\hspace*{10pt} In our experiments, we use a single crystal HPHT diamond with dimensions of 2 mm $\times$ 2 mm $\times$ 0.7 mm, having a $\{100\}$ front facet and a NV concentration of nearly 2.6 ppm. The diamond crystal coordinate system ($x,y,z$) having the unit vectors $\{\bm{e}_x,\bm{e}_y,\bm{e}_z\}$ is defined such that $\bm{e}_x\parallel[100],\;\bm{e}_y\parallel[010]\;\text{and}\;\bm{e}_z\parallel[001]$ where $[100],\;[010],\;\text{and}\;[001]$ are the diamond lattice vectors (see Fig. \ref{fig:heatmap}a). Due to the tetrahedral nature of the diamond lattice, there are four possible orientations of the NV centers (see Fig. \ref{fig:heatmap}a). The unit vectors $\bm{e}_{i}$, $i\in\{\kappa,\chi,\varphi,\lambda\}$, corresponding to these NV orientations are $\bm{e}_{\kappa}=(\sqrt{1/3},\sqrt{1/3},\sqrt{1/3})^T\parallel[111]$, $\bm{e}_{\chi}=(-\sqrt{1/3},-\sqrt{1/3},\sqrt{1/3})^T\parallel[\Bar{1}\Bar{1}1]$, $\bm{e}_{\varphi}=(-\sqrt{1/3},\sqrt{1/3},-\sqrt{1/3})^T\parallel[\Bar{1}1\Bar{1}]$ and $\bm{e}_{\lambda}=(\sqrt{1/3},-\sqrt{1/3}, -\sqrt{1/3})^T\parallel[1\Bar{1}\Bar{1}]$. \\
\hspace*{10pt} Upon the application of a magnetic field $\vb*{B}$, each NV orientation may experience a different projection of the magnetic field, leading to varied Zeeman splitting. Fig. \ref{fig:heatmap}b illustrates one such case where $\vb*{B}$ is oriented such that each NV center experiences a different projection along the NV axis (see Fig. \ref{fig:heatmap}c), thus resulting in an ODMR spectrum with 8 dips. The ground state spin Hamiltonian for each of the four NV orientations $i\in\{\kappa,\chi,\varphi,\lambda\}$ under the influence of $\vb*{B}$ reads as \cite{Tetienne:2012, Rondin:2014, Schloss_2018}:
\begin{equation}
    \label{Eq:Hamiltonian1}
    \mathcal{\hat{H}}^{i}=D (\hat{S}_z^i)^2 +\gamma \vb*{B}\cdot\hat{\bm{S}}^i
\end{equation} 
where $D=2.87$ GHz is the zero-field splitting (ZFS), $\gamma = 28$ MHz/mT is the gyromagnetic ratio of the NV center, $\vb*{\hat{S}}^i = (\hat{S}_x^i, \hat{S}_y^i, \hat{S}_z^i)^T$ represents the dimensionless NV electronic spin-1 operators in the NV$_i$ coordinate system, and $\vb*{B}$ is the external magnetic field vector. Here, we do not consider NV hyperfine interactions, since our ODMR measurements are not hyperfine resolved due to the large concentration of P1 spin impurities.\\
\hspace*{10pt} In this study, we focus only on small static magnetic fields ($<$4 mT). Thus, to first order, the transverse components of $\vb*{B}$ can be ignored \cite{Tetienne:2012, Rondin:2014}. This reduces Eq. (\ref{Eq:Hamiltonian1}) to:
\begin{equation}
    \label{Eq:Hamiltonian2}
    \mathcal{\hat{H}}^{i}=D (\hat{S}_z^i)^2 +\gamma B_{i}\hat{S}^i_z
\end{equation} 
where $B_{i}=\vb*{B}\cdot\vb*{e}_{i}$ is the projection of the magnetic field $\vb*{B}$ along the NV$_i$ orientation. The transition frequencies $\{f_{i-},f_{i+}\}$ for a given NV$_i$ orientation are:
\begin{equation}
    \label{Eq:splitting1}
    f_{i\pm}=D\pm\gamma B_{i}
\end{equation}
Thus, the number of dips observed in the ODMR spectrum depends on the orientation of the magnetic field with respect to the four crystallographic orientations of the NV center. In principle, all the NVs can have a different magnetic field projection for an arbitrary orientation of the magnetic field, leading to eight dips in the ODMR spectrum  (Figs. \ref{fig:heatmap}(b) and \ref{fig:heatmap}(c)). However, the number of dips observed in the ODMR spectrum is reduced for certain directions of the magnetic field owing to symmetry considerations. Fig. \ref{fig:exp_fig}(c) shows a heat-map simulation of the various possible number of ODMR dips, which arise from different polar ($\theta$) and azimuthal ($\phi$) angles of the magnetic field $\vb*{B}$ in the diamond crystal coordinate system. We discuss this in more detail in section \ref{section:results}. \\
\hspace*{10pt} Fig. \ref{fig:exp_fig} shows a schematic of the experimental setup used to systematically study the above relation between $\vb*{B}$ and the NV orientations, leading to various different ODMR spectra. The core unit of the experimental setup consists of a 30 mm cage cube (Thorlabs C4W). The cube mount helps in easy alignment of the optics, and also provides portability and stability to the setup. The single mode fiber cable from the fiber-coupled laser is connected to a fiber collimator (0.3 NA) mounted on the cube to collimate and guide the linearly polarized laser beam towards a rectangular dichroic mirror (550 nm cut-on, 25 mm $\times$ 36 mm lateral dimensions). The green laser beam gets reflected by the dichroic mirror and passes through a microscope air objective (0.75 NA, 100\ba{x})  which focuses it onto the diamond sample. Thus, optical excitation of the NV centers in diamond takes place, thereby causing emission of red photoluminescence (PL). The objective lens also collects a small fraction of the emitted red PL and directs it towards the dichroic mirror. The PL passes through the dichroic mirror, which is then filtered through a long-pass filter (650 nm cut-off). The filtered beam is then focused onto a photodiode sensor (Thorlabs PDA100A2) using a condenser lens (Thorlabs ACL25416U-B). The photodiode converts incident PL into an analog voltage, which is then acquired by the data acquisition hardware (NI DAQ 6002). Finally, the DAQ signal is further processed and displayed using LabVIEW software on the computer. \\
\hspace*{10pt} In our setup, the diamond sample is mounted external to the optical setup just described. It is glued to a printed-circuit board (PCB), which delivers the MW radiation to the NV centers via a 20 $\mu$m copper wire placed on top of the diamond surface. The MW signal is generated by a signal generator (SG), which is then amplified by a MW amplifier before being transmitted to the PCB. For minor adjustments of the diamond relative to the green laser beam spot, the PCB with the diamond is mounted on a micro-positioning setup. We use two circular electromagnetic coils for achieving control over the direction and magnitude of the DC magnetic field generated in relation to all the NV orientations in the diamond. As illustrated in Fig. \ref{fig:exp_fig}, the \ba{X} component of the magnetic field generated by both the coils will be zero as the axes of the coils are aligned along the \ba{Y} and \ba{Z} axes of the laboratory coordinate system. Thus, by varying the DC current through the Coil 1 and Coil 2, one can achieve control over the polar angle $\theta$ of $\vb*{B}$ only. Hence, in some cases when the control over the azimuthal angle $\phi$ is required, the diamond is rotated about the \ba{Z} axis of the laboratory coordinate system in order to obtain the desired spatial orientation of $\vb*{B}$ relative to the crystallographic axes for conducting the experiment. 
\begin{figure}[h]
\centering
\includegraphics[width=1\textwidth]{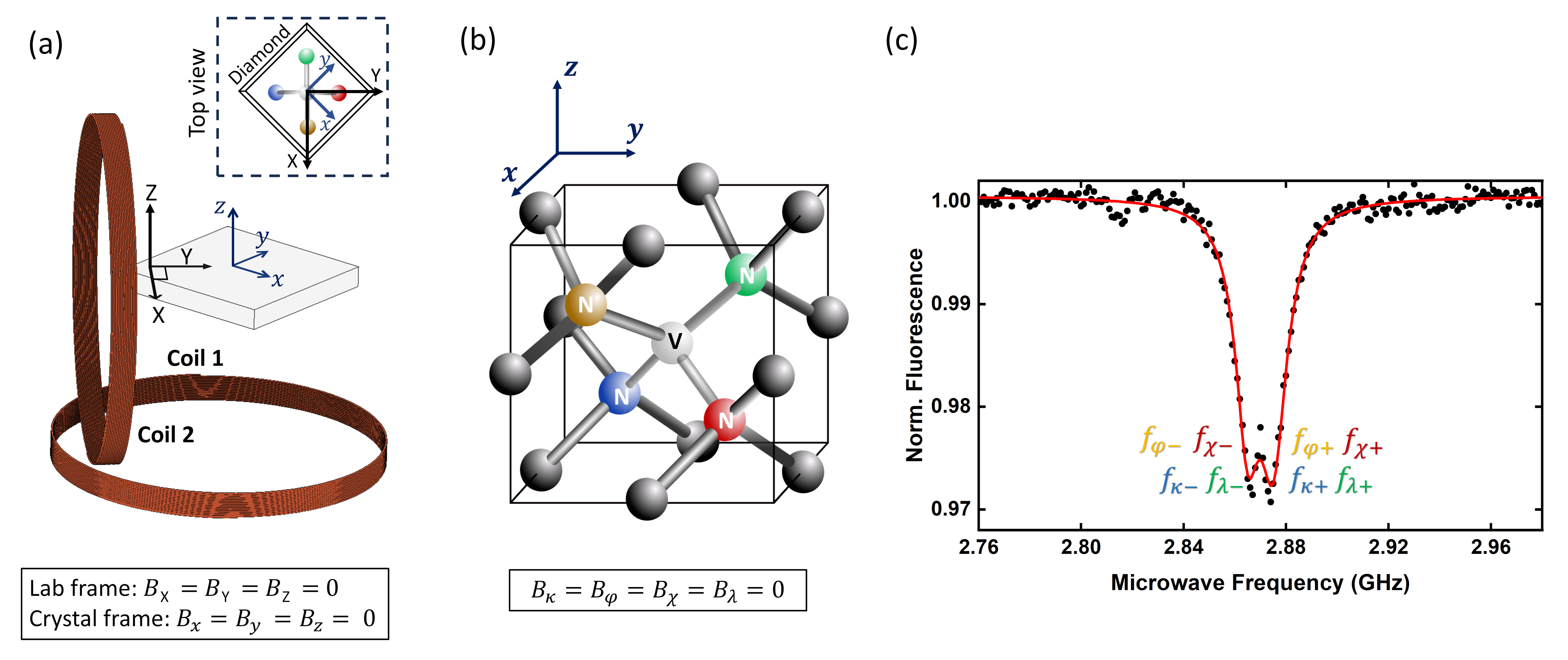}
\caption{(a) Illustration of the experimental setup for Case 1, where the current flowing through both the coils is zero, i.e., $|\vb*{B}|$ = 0. (b)  The projection of the magnetic field onto the NV orientations is zero. 
(c) A single dip corresponding to the ZFS is observed in the ODMR spectrum. However, a splitting within this ZFS dip is also observed and a two-peak Lorentzian fit is performed by considering the intrinsic effective field present in the diamond lattice.}
\label{fig:case_1}
\end{figure}
\section{Results and discussion}
\label{section:results}
As discussed in the previous section, the angle \(\theta\) of the magnetic field vector $\vb*{B} = (B_x,B_y,B_z)^T$ in the diamond crystal coordinate system can be varied by adjusting the currents in the coils and the angle \(\phi\) can be varied by rotating the diamond about the \ba{Z} axis of the laboratory
coordinate system. The vector components of the applied field 
\(\vb*{B}\) can therefore be written as $B_x=|\vb*{B}|\,\sin{\theta}\cos{\phi},\;B_y=|\vb*{B}|\,\sin{\theta}\sin{\phi},\ $ and $B_z=|\vb*{B}|\,\cos{\theta}$. By constructing a \(4 \times 3\) transformation matrix $\vb*{T}$ from the unit vectors corresponding to the four NV orientations, 
\(\vb*{e}_i\) for \(i \in \{\kappa, \chi, \varphi, \lambda\}\), we obtain a simplified expression for $\mathcal{B}_{NV}=(B_\kappa,B_\chi,B_\varphi,B_\lambda)^T$. The matrix equation is
\begin{subequations}
\label{fig:projec}
    \begin{align}
        \begin{bmatrix}
             B_\kappa \\
            B_\chi \\
             B_\varphi \\
             B_\lambda
        \end{bmatrix} &= \vb*{T}\begin{bmatrix}
             B_x \\
            B_y \\
             B_z
        \end{bmatrix}, \label{Eq:projec}\\
        \intertext{where}
        \vb*{T} &= \frac{1}{\sqrt{3}}
        \begin{bmatrix}
             1 & 1 & 1 \\
            -1 & -1 & 1 \\
            -1 & 1 & -1 \\
             1 & -1 & -1
        \end{bmatrix}.
    \end{align}
\end{subequations}
Eq. \ref{Eq:splitting1} shows that the projections $B_i$ are directly related to the dips observed in the ODMR spectrum.  However, owing to the tetrahedral symmetry of the diamond lattice, these projections must satisfy the relation
\begin{equation}
\label{Eq:projec_relation}
    B_{\kappa} + B_{\chi} + B_{\varphi} + B_{\lambda} = 0.
\end{equation}
Thus, Eqs. \ref{Eq:projec} and \ref{Eq:projec_relation} together impose sign and magnitude constraints on the possible combinations of \(B_i\) required for observing a particular number of dips in the ODMR spectrum. This, in turn, guides us in determining the orientation of the magnetic field vector relative to the NV axes. \\
\hspace*{10pt}Below, we outline one representative set of conditions on the Cartesian components \((B_x, B_y, B_z)\) for each of the eight possible cases (see Fig. \ref{fig:exp_fig}(c)) and also experimentally obtain the corresponding ODMR spectrum by generating the required magnetic field vector $\vb*{B}$ in our setup. Owing to the eight-fold symmetry in the heat-map, we restrict our theoretical analysis of \((B_x, B_y, B_z)\) and the corresponding experimental observation of the ODMR spectrum to the first quadrant of the diamond crystal coordinate system ($x,y,z$). For an ODMR spectrum having a certain number of dips, an exhaustive list of associated magnetic field vectors \((B_x, B_y, B_z)\) is described in supplementary information.
\begin{figure}[h]
\centering
\includegraphics[width=1\textwidth]{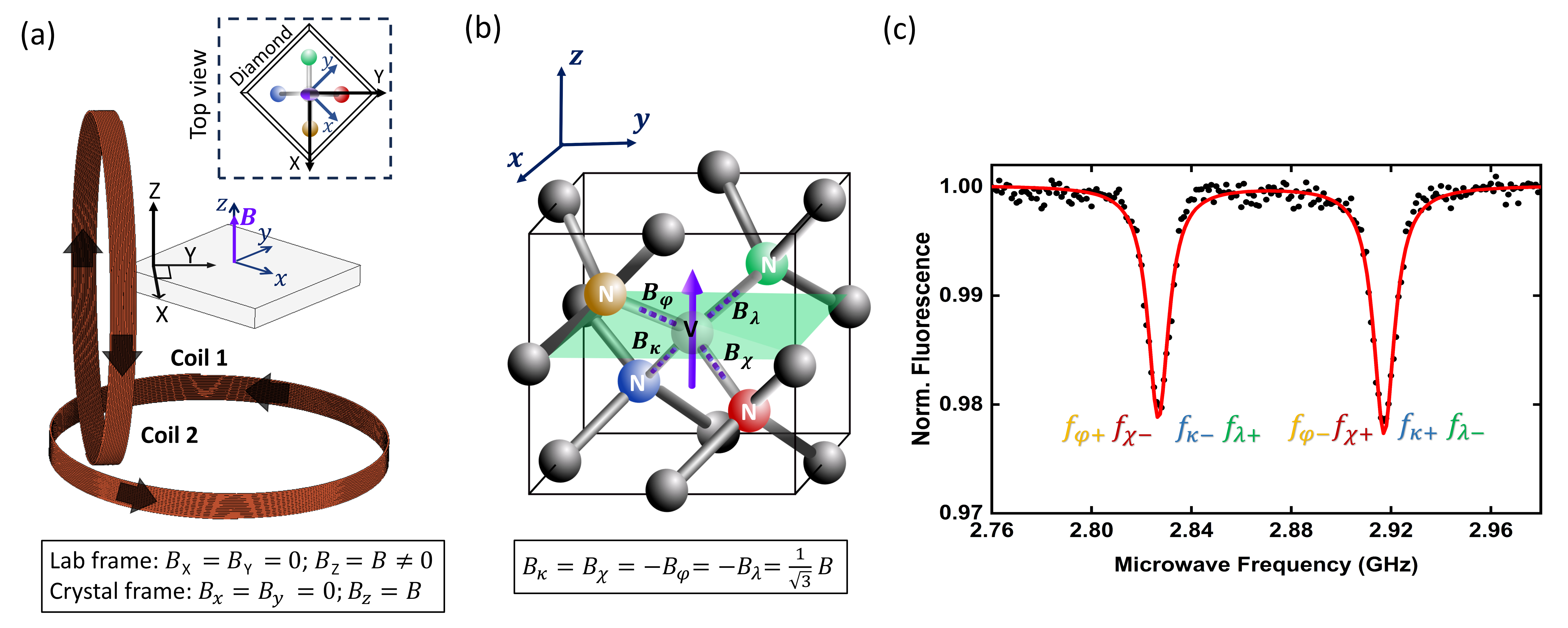}
\caption{(a) Illustration of the experimental setup for Case 2, where the magnetic field is applied along the \ba{Z}-axis using Coil 1 only. (b) The applied field $\vb*{B}$ is perpendicular to the green-shaded plane. Due to the tetrahedral symmetry of diamond lattice, the field projections on all four NV orientations share the same magnitude equal to $\frac{B}{\sqrt{3}}$. (c) Consequently, the Zeeman-split resonances of all NV orientations overlap, producing only two distinct dips in the ODMR spectrum.}
\label{fig:case_2}
\end{figure}
\subsection{\textbf{Case 1: Condition for one dip}}
\label{subsection:case1}
\begingroup
When there is no applied magnetic field, i.e., $\vb*{B}$ = 0, the projection of the magnetic field $B_i$ on each NV orientation is zero (see Fig. \ref{fig:case_1}(b)). Thus, only a single dip is observed in the ODMR spectrum at $\sim$2.87 GHz corresponding to ZFS (see Fig. \ref{fig:case_1}(c)). However, in our experiment, despite the absence of $\vb*{B}$, a splitting of approximately 10 MHz is observed. It is attributed to a local intrinsic effective field arising from electric and strain fields present in the diamond lattice \cite{Mittiga:2018, Kolbl:2019, Kumar:2023}. This effect is not considered in this study, since it is not relevant to our investigations. Hence, in the rest of the cases, we have fitted the dip arising due to ZFS with a single Lorentzian. 
\subsection{\textbf{Case 2: Condition for two dips}}
\label{subsection:case2}
If we apply an external magnetic field $\vb*{B}$ such that the magnitude of the projection is the same for all the four NV center orientations (NV$_i$),  then the resulting Zeeman splitting ($\Delta f_i$) will also be the same for all orientations (see Fig. \ref{fig:case_2}). This configuration of $\vb*{B}$ leads to two distinct dips in the ODMR spectrum owing to the overlap of Zeeman-split resonances of all the NV center families (see Fig. \ref{fig:case_2}(c)). To achieve this, the magnetic field can be applied along one of the six equivalent \(\langle100\rangle\) crystallographic directions. In our experimental setup, this condition is being met by aligning $\vb*{B}$  along the $z$-axis of the diamond crystal, as illustrated in the Fig. \ref{fig:case_2}(a). Then:
    \begin{equation}
    \label{Eq:twodips_cond}
    B_{\kappa} = B_{\chi} = -B_{\varphi} = -B_{\lambda} = \frac{1}{\sqrt{3}}B.
    \end{equation}
The resulting ODMR spectrum with two dips is shown in Fig. \ref{fig:case_2}(c).
\begin{figure}[h]
\centering
\includegraphics[width=1\textwidth]{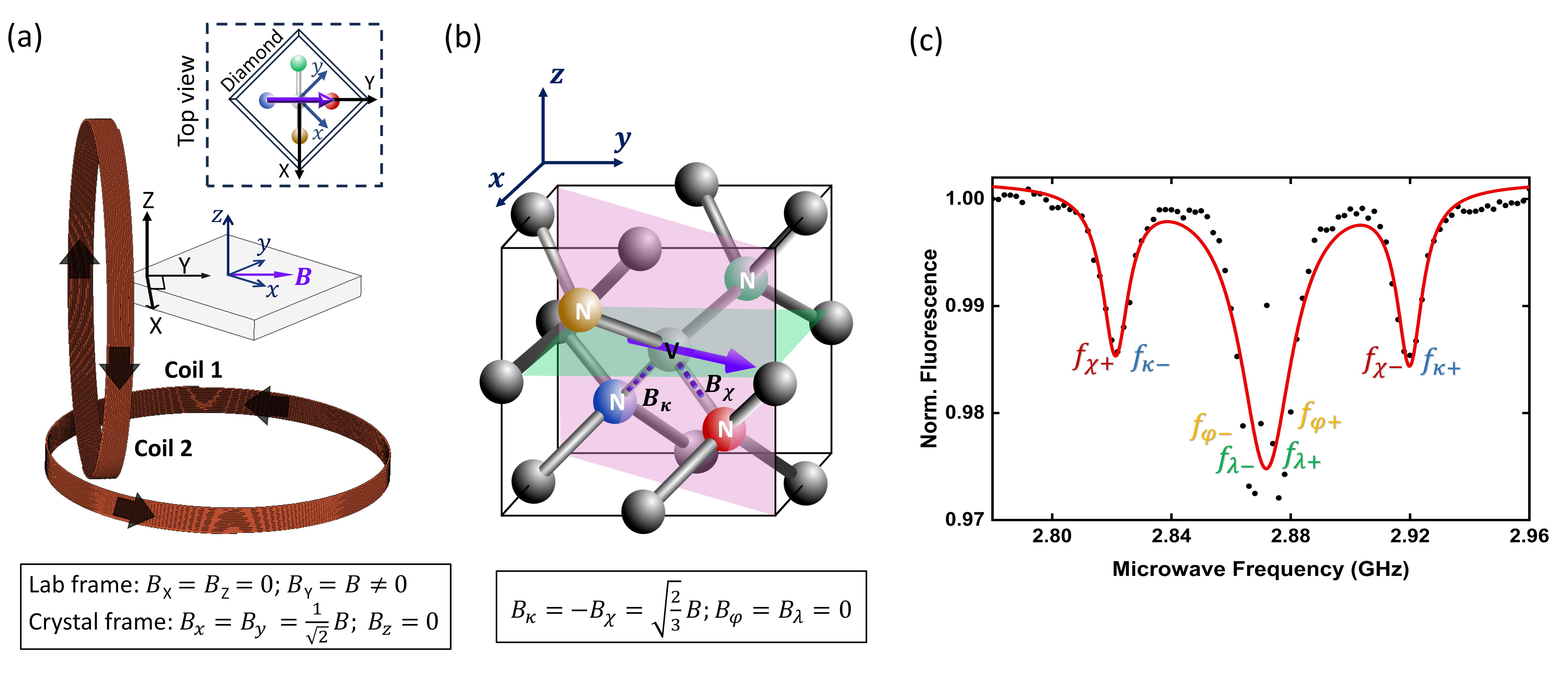}
\caption{
(a) Schematic of the experimental setup for Case 3, where the magnetic field is applied along the \(Y\)-axis using Coil 2. The resulting field orientation satisfies \(B_x = B_y\), as also illustrated in the inset. 
(b) The applied field lies along the intersection of the green and purple planes, ensuring that the projections onto NV\(_{\varphi}\) and NV\(_{\lambda}\) are zero, while NV\(_{\kappa}\) and NV\(_{\chi}\) experience equal and opposite projections. 
(c) Consequently, three distinct dips appear in the ODMR spectrum due to the overlapping resonances of NV\(_{\varphi}\) and NV\(_{\lambda}\).
}
\label{fig:case_3}
\end{figure}
\begin{figure}[h]
\centering
\includegraphics[width=1\textwidth]{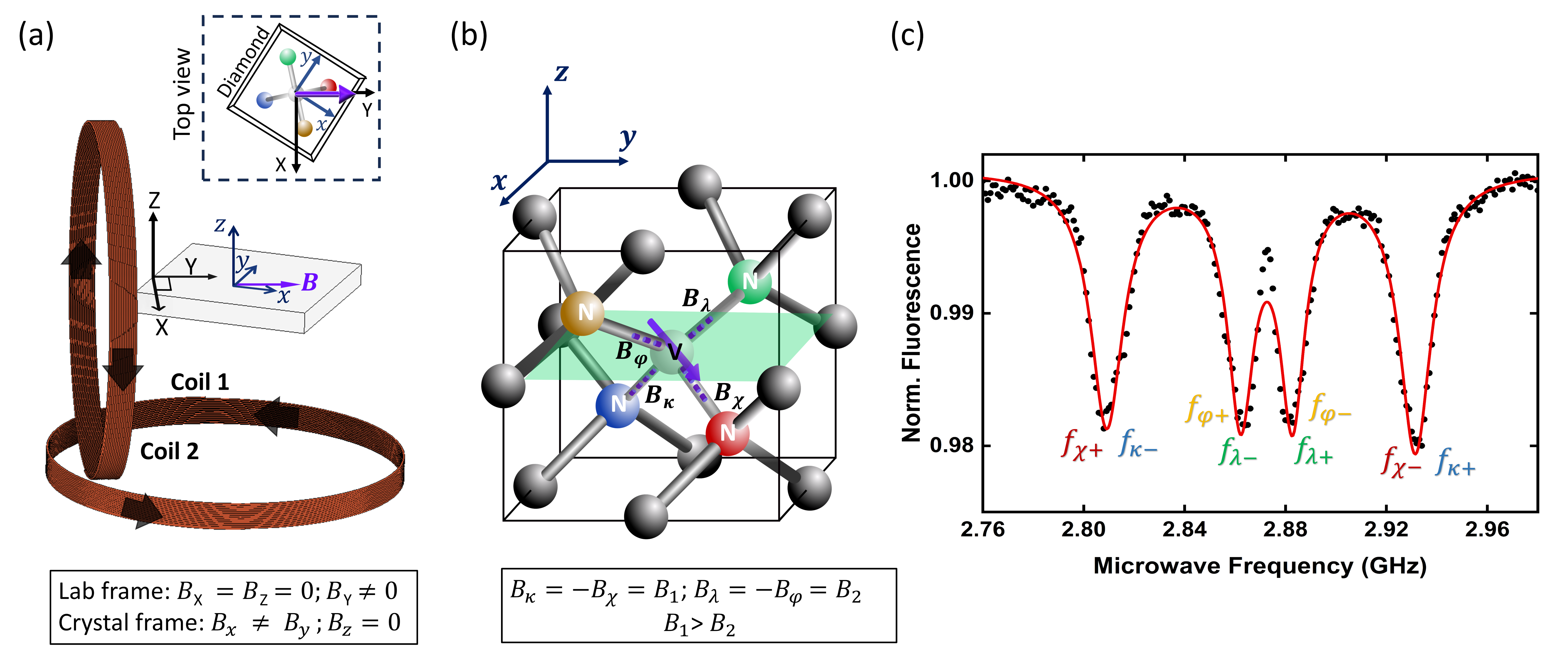}
\caption{
(a) Schematic of the experimental setup for Case 4a, where the magnetic field satisfies the relations \(B_x \neq B_y \neq 0\) and \(B_z = 0\). We note that the field is applied using Coil 2 only. The inset illustrates the orientation of the diamond crystal relative to the laboratory coordinate system, ensuring that the required condition \(B_x \neq B_y \neq 0 \) is met. 
(b) In contrast to Case 3, the configuration of \(\vb*{B}\) makes equal magnitude projections for the other NV pair (NV\(_\varphi\) and NV\(_\lambda\)) as well. 
(c) Four ODMR dips are observed in this case, with each dip pair corresponding to one of the two NV pairs, as indicated in the spectrum.
}
\label{fig:case_4a}
\end{figure}
\subsection{\textbf{Case 3: Condition for three dips}}
\label{subsection:case3}
This case arises when $\vb*{B}$ is applied along one of the twelve equivalent \(\langle110\rangle\) crystallographic directions. Then, one NV pair (say NV$_{\kappa}$ and NV$_{\chi}$) has projections of equal magnitude while the other pair (say NV$_{\varphi}$ and NV$_{\lambda}$) has zero projection. As shown in Fig. \ref{fig:case_3}(b), the field is normal to the plane containing two particular NV orientations, NV$_{\varphi}$ and NV$_{\lambda}$. Consequently, the projections onto these two NV orientations will be zero, and the resulting ODMR spectrum of these two NV orientations will have a single dip at the ZFS frequency. In contrast, $\vb*{B}$ lies within the plane containing the other two NV orientations, NV$_{\kappa}$ and NV$_{\chi}$. Hence, both the projections will be equal in magnitude, but opposite in sign leading to a finite Zeeman splitting $\Delta f=2\gamma B\cos{(35.25^{\circ}})$. Thus:
\begin{equation}
    \label{Eq:threedips_cond}
    B_{\kappa} = -B_{\chi} = \sqrt{\frac{2}{3}}B; \quad B_{\varphi} = B_{\lambda} = 0.
    \end{equation}
This configuration results in three distinct dips in the ODMR spectrum: the central dip corresponds to the NV$_{\varphi}$ and NV$_{\lambda}$  orientations (without Zeeman splitting), while the outer two dips correspond to the N$V_{\kappa}$ and NV$_{\chi}$ orientations (with Zeeman splitting).
The resulting three-dip ODMR spectrum is shown in Fig. \ref{fig:case_3}(c). 
\subsection{\textbf{Case 4: Condition for four dips}}
\label{subsection:case4}
There are two possible scenarios that can produce four dips in the ODMR spectrum:
\begin{enumerate}
    \item \textbf{Case 4a}: Four dips in the ODMR spectrum can arise when one of the three components of the magnetic field \(\vb*{B}\) in the crystal coordinate system is zero and the remaining two have unequal and non-zero magnitudes. For example, consider \(\vb*{B}\) oriented along $(B_x,B_y,0)^T$ direction, where $B_x$ and $B_y$ can take any values satisfying the condition $B_x\neq B_y \neq 0$. Then, the magnetic field \(\vb*{B}\) results in identical Zeeman splittings for the NV$_{\kappa}$ and NV$_{\chi}$, and the NV$_{\varphi}$ and NV$_{\lambda}$ orientations, respectively. Thus, 
        \begin{equation}
        \label{Eq:fourdips_a_cond}
        B_{\kappa} = -B_{\chi} = B_1; B_{\lambda} = -B_{\varphi} = B_2.
        \end{equation}
    In the Fig. \ref{fig:case_4a}(a), since \(\vb*{B}\) is only slightly misaligned from the [110] crystallographic direction,  
    the inner pair corresponds to the NV$_{\varphi}$ and NV$_{\lambda}$ orientations and the outer pair corresponds to the NV$_{\kappa}$ and NV$_{\chi}$ orientations. We note that there are 24 possible magnetic field vectors that produce the same ODMR spectrum. For this case, only the coil aligned along the \ba{Y}-axis is used for generating the desired magnetic field (see Fig. \ref{fig:case_4a}(a)).
    \item \textbf{Case 4b}: Here, $\vb*{B}$ is applied along one of the eight equivalent ⟨111⟩ crystallographic directions. Without loss of generality, let \(\vb*{B}\) point along NV\(_\kappa\). 
    Thus, due to the tetrahedral symmetry, the magnitudes of the projections will be identical for NV\(_\varphi\), 
    NV\(_\chi\) and NV\(_\lambda\) orientations. This results in:
        \begin{equation}
        \label{Eq:fourdips_b_cond}
        \frac{1}{3}B_{\kappa}=-B_{\chi} = -B_{\varphi} = -B_{\lambda} = \frac{1}{3}B.
        \end{equation}
     Figure~\ref{fig:case_4b}(b) illustrates this configuration leading to 
     four dips in the ODMR spectrum. The outermost pair of dips corresponds to the NV$_{\kappa}$, along which $\vb*{B}$ is aligned. The inner pair of dips represents the contribution from the remaining three NV orientation classes (Fig. \ref{fig:case_4b}(c)). In this case, both the coils are used simultaneously and the current through the coils adjusted to ensure that all the three components of the vector $\vb*{B}$ have equal magnitude in the crystal coordinate system (see Fig. \ref{fig:case_4b}(a)).
\end{enumerate}
\begin{figure}
\centering
\includegraphics[width=1\textwidth]{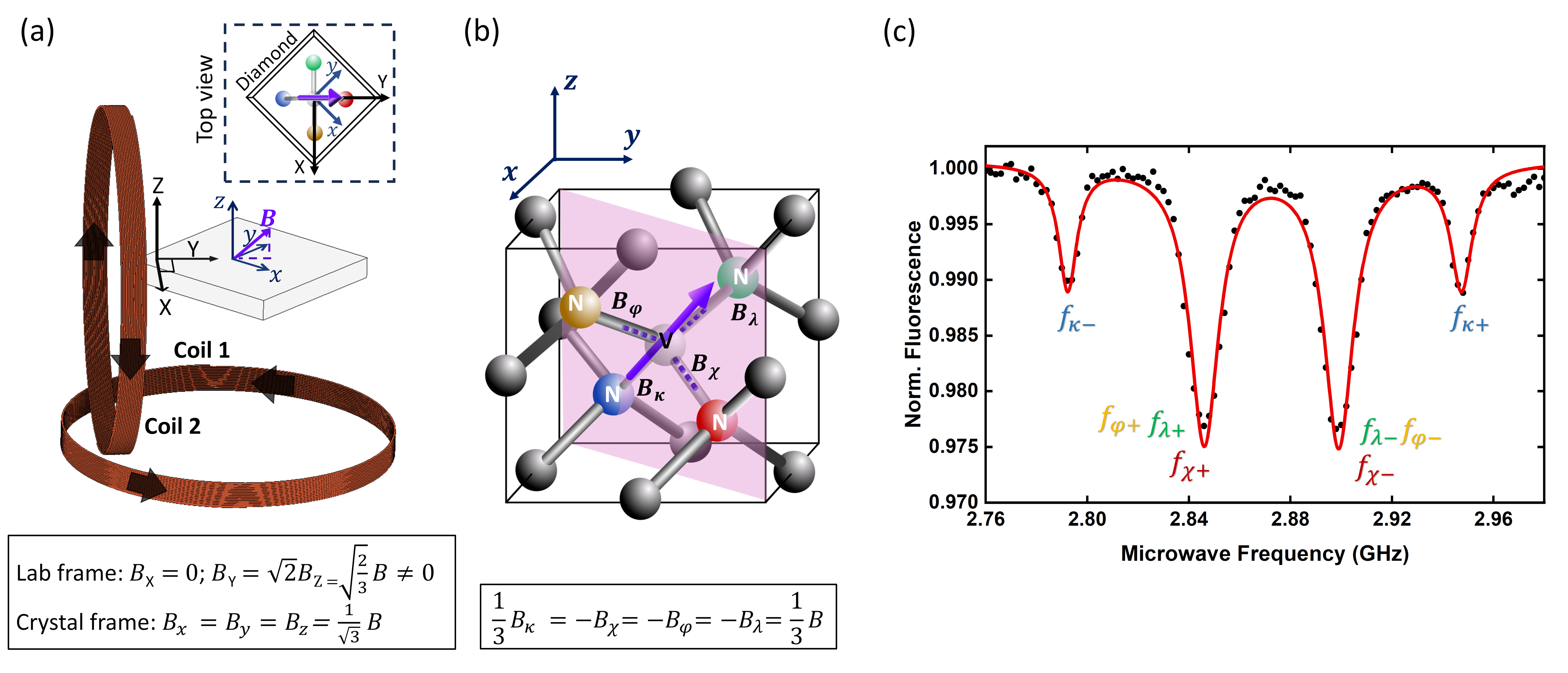}
\caption{
(a) Schematic of the experimental setup for Case~4b, where the magnetic field \(\vb*{B}\) is aligned along an NV axis, thereby satisfying the relation \(B_x = B_y = B_z \neq 0\). This configuration is achieved by driving current through both Coil~1 and Coil~2 simultaneously in order to generate the required \(\vb*{B}\). 
(b) \(\vb*{B}\) is directed along the NV\(_\kappa\) axis. Due to the tetrahedral symmetry of the diamond lattice, it projects equally onto the remaining three NV orientations. 
(c) The ODMR spectrum exhibits four dips, with the outermost dips corresponding to the NV\(_\kappa\) orientation, along which \(\vb*{B}\) is applied. 
}

\label{fig:case_4b}
\end{figure}
\subsection{\textbf{Case 5: Condition for 5 dips}}
\label{subsection:case5}
In contrast to Case 3, this case occurs when $\vb*{B}$ is perpendicular to only one (say NV\(_\chi\)) of the four NV orientations leading to a corresponding dip in the ODMR spectrum at the ZFS frequency and the projections on the remaining three NV orientations contain one pair (say NV\(_\varphi\) and NV\(_\lambda\)) of the same sign and magnitude. Then, from Eq. \ref{Eq:projec_relation}, we get
    \begin{equation}
    \label{Eq:fivedips_cond}
    \frac{1}{2}B_{\kappa}=-B_{\varphi}=-B_{\lambda}=\frac{1}{2}\sqrt{\frac{8}{9}}B; B_{\chi} = 0.
    \end{equation}
Fig. \ref{fig:case_5}(b) illustrates the magnetic field configuration in the diamond lattice satisfying the above conditions. We obtain this field experimentally by using both the coils such that the Coil 1's field strength is roughly equal to $\sqrt{2}$ times that of Coil 2 [Fig. \ref{fig:case_5}(a)]. The diamond crystal coordinate system maintains the same orientation as described in Case 4b, ensuring that the magnetic field projects as required onto all four NV orientations as depicted in Fig. \ref{fig:case_5}(b). This results in the expected five-dip ODMR spectrum shown in Fig. \ref{fig:case_5}(c). We note that there are 24 possible magnetic field vectors that can produce the same ODMR
spectrum having five dips.
\begin{figure}
\centering
\includegraphics[width=1\textwidth]{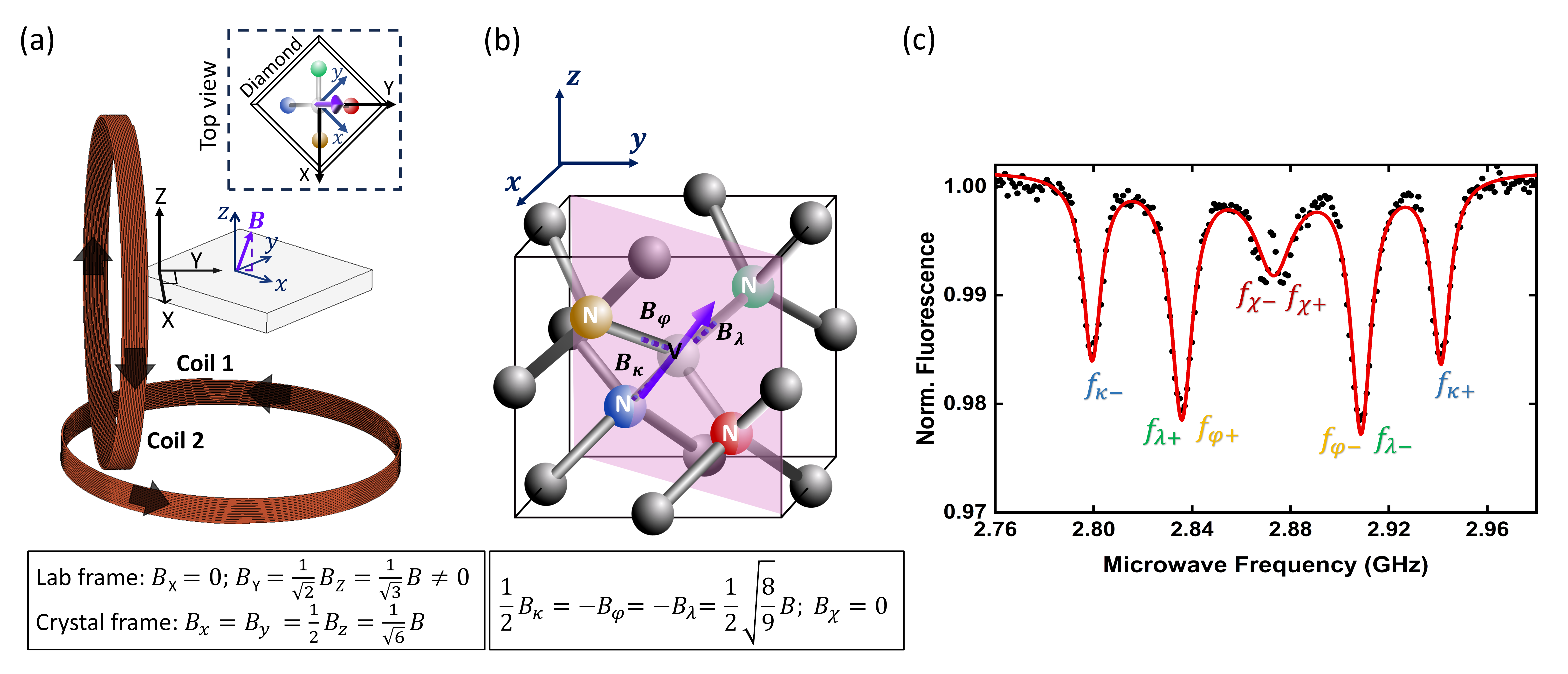}
\caption{
(a) Schematic of the experimental setup for Case~5, where both coils are used to generate the magnetic field \(\vb*{B}\). The magnetic fields from the two coils are adjusted such that \(B_x = B_y \neq 0\) and \(B_z = 2B_x\).
(b) In this configuration, \(\vb*{B}\) lies in the plane containing NV\(_\kappa\) and NV\(_\chi\). The projection onto NV\(_\chi\) is zero, while NV\(_\kappa\) experiences the maximum Zeeman splitting. The projections onto the other NV pair (NV\(_\lambda\) and NV\(_\varphi\)) are equal in magnitude.  
(c) The resulting ODMR spectrum exhibits five dips, confirming the expected NV resonance shifts under the applied field.
}
\label{fig:case_5}
\end{figure}
\subsection{\textbf{Case 6: Condition for 6 dips}}
\label{subsection:case6}
This case features one NV pair (say NV\(_\lambda\) and NV\(_\varphi\)) with equal non-zero projections, while the other two NVs have unequal non-zero projections whose magnitudes are different from that of the former. Then,
    \begin{equation}
    \label{Eq:sixdips_cond}
    B_{\kappa} = B_1 ; B_{\chi} = B_2 ; B_{\varphi} = B_{\lambda} = -B_3 = -\frac{1}{2}(B_1+B_2).
    \end{equation}
The magnetic field \(\vb*{B}\) depicted in Fig.~\ref{fig:case_6}(b) satisfies the above conditions. The field lies within the plane spanned by NV\(_\kappa\) and NV\(_\chi\), with its components in the diamond crystal coordinate system satisfying \(B_x = B_y\). While performing the experiment, we ensure that the current flowing through the Coil 2 is such that the resulting \(B_z\) does not correspond to the cases considered earlier. The orientation of the diamond crystal is as shown in the Fig.~\ref{fig:case_6}(a), ensuring that the applied field \(\vb*{B}\) leads to the expected magnetic field projections on the NV orientations. The experimental setup and the corresponding six-dip ODMR spectrum are  shown in Figs.~\ref{fig:case_6}(a) and ~\ref{fig:case_6}(c). We note that given the measured ODMR spectrum with six dips, there are 24 possible magnetic field vectors that can reproduce it. 
\begin{figure}
\centering
\includegraphics[width=1\textwidth]{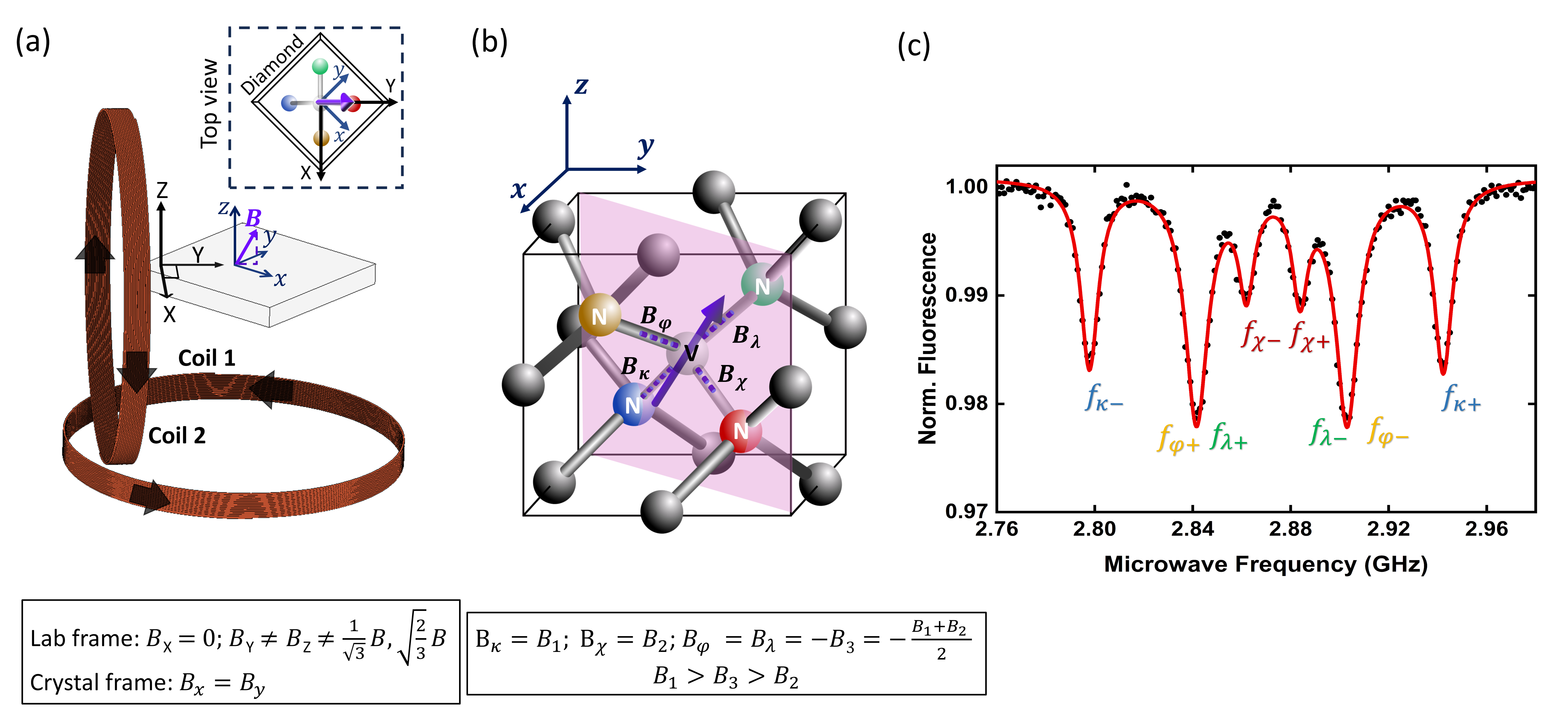}
\caption{
(a) Schematic of the experimental setup for Case 6, where both coils are used to generate the magnetic field \(\vb*{B}\). 
(b) In this configuration, \(\vb*{B}\) lies within the plane containing the NV orientations NV\(_\kappa\) and NV\(_\chi\), but with different values of projection on these two NV orientations. The projections onto the other NV pair (NV\(_\lambda\) and NV\(_\varphi\)) are equal.  
(c) This configuration results in an ODMR spectrum with six dips.
}
\label{fig:case_6}
\end{figure}
\subsection{\textbf{Case 7: Condition for 7 dips}}
\label{subsection:case7}
\begin{figure}
\centering
\includegraphics[width=1\textwidth]{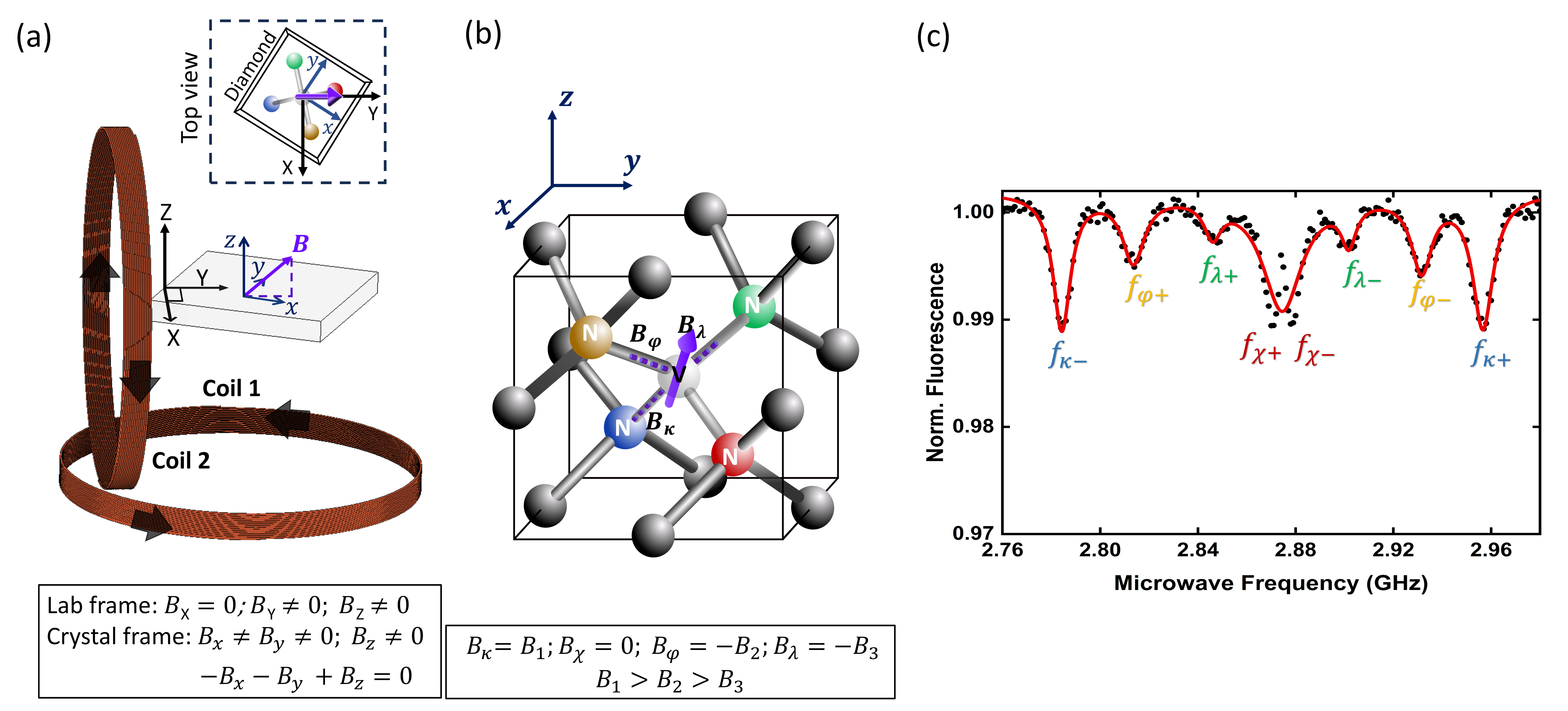}
\caption{
(a) Schematic of the experimental setup for Case~7, where both coils are used to generate the magnetic field \(\vb*{B}\). The field components are adjusted so that \(B_x \neq B_y\) while satisfying the constraint \(B_{\chi}=- B_x - B_y +B_z = 0\).  
(b) In this configuration, three of the four NV orientations experience distinct magnetic field projections, while the fourth projection is zero.  
(c) The resulting ODMR spectrum features seven dips, with three pairs of dips corresponding to three NV orientations, and the dip at ZFS is due to the projection being zero on the fourth NV orientation.
}
\label{fig:case_7}
\end{figure}
\begin{figure}
\centering
\includegraphics[width=1.0\textwidth]{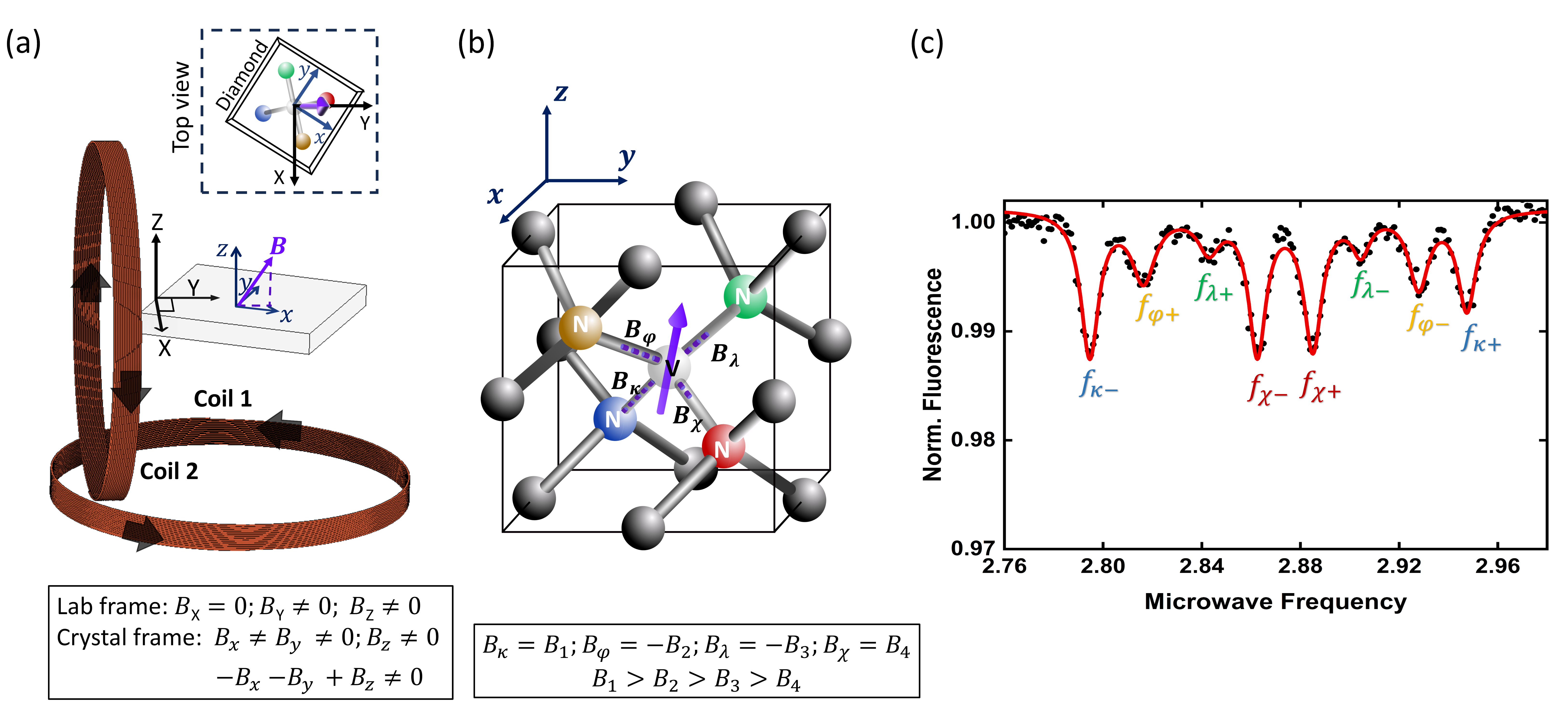}
\caption{
(a) Schematic of the experimental setup for Case~8, where both coils are used to generate the magnetic field \(\vb*{B}\). Starting from the current configuration of Case 7, the current values are slightly perturbed such that the condition \(-B_x - B_y + B_z \neq 0\) is satisfied. The inset provides a top view of the diamond crystal orientation relative to the lab frame.  
(b) In this configuration, the magnitude of the magnetic field projection is distinct for all four NV orientations.  
(c) The resulting ODMR spectrum exhibits eight distinct dips or four pairs of dips, with each pair corresponding to a specific NV orientation.
}
\label{fig:case_8}
\end{figure}
This case arises when $\vb*{B}$ is perpendicular to only one NV orientation (say NV\(_\chi\)) and has non-zero projection of different magnitudes on the remaining three NV orientations. This case can also be understood as arising from the Case 5 by modifying the condition from \(B_x = B_y\neq0\) to \(B_x \neq B_y \neq 0\). Then, 
    \begin{equation}
    \label{Eq:sevendips_cond}
    B_{\chi} = 0;\quad |B_{\varphi}| \neq |B_{\kappa}| \neq |B_{\lambda}| \neq 0.
    \end{equation}
Fig. \ref{fig:case_7}(b) illustrates the magnetic field vector $\vb*{B}$ satisfying the above conditions. To experimentally obtain this field, both coils are used such that \(B_Y \neq B_Z \neq 0\). As shown in the Fig.~\ref{fig:case_7}(a), the diamond crystal orientation is slightly rotated such that $\vb*{B}$ does not lie in the plane spanned by NV$_\kappa$ and NV$_\chi$ anymore (similar to Case 4a). 
The seven-dip ODMR spectrum is shown in Fig.~\ref{fig:case_7}(c). 
We note that there are 24 possible magnetic field vectors that can produce the same ODMR spectrum with seven dips.    
\subsection{\textbf{Case 8: Condition for 8 dips}}
\label{subsection:case8}
Finally, the maximum number of dips occur when all four NV orientations experience distinct, non-zero projections:
    \begin{equation}
    \label{Eq:eightdips_cond}
    |B_{\kappa}| \neq |B_{\lambda}| \neq |B_{\varphi}| \neq |B_{\chi}| \neq 0.
    \end{equation}
The magnetic field $\vb*{B}$ illustrated in Fig. \ref{fig:case_8}(b) satisfies the above required conditions. The diamond crystal orientation is as shown in Fig. \ref{fig:case_8}(a). Both the Coil 1 and Coil 2 are used to generate the desired field $\vb*{B}$. 
The eight dip ODMR spectrum is shown in Fig. \ref{fig:case_8}(c), where each pair of dips corresponds to one NV orientation. We note that there are 48 possible magnetic field vectors that can produce the same ODMR spectrum with eight dips. 

\section{Conclusion}
\label{section:conclusion}
In conclusion, we performed a comprehensive study of the various possible low-field CW-ODMR spectra with an ensemble of NV centers by exploring the geometric relationship between the applied magnetic field and the NV intrinsic quantization axes. Our theoretical model and measurement data uncover an intricate geometrical dependence of the ODMR spectra, which can find widespread utility in vector magnetometry applications. Our studies can aid vector magnetic sensing in applications where a low bias field is desired to avoid unwanted magnetization effects. 
\section*{ACKNOWLEDGMENTS}
P P gratefully acknowledges financial support from I-HUB Quantum Technology Foundation Grant Number I-HUB/SPOKE/2023-24/003 and STARS MoE Grant Number MoE-STARS/STARS1/662. S K acknowledges the CSIR funding agency Award Number 09/1020(0202)/2020-EMR-I for providing SRF fellowship. P D acknowledges Roshin G Vallatt and Sakshi Dwivedy for their assistance in the experimental measurements. 
\section*{AUTHOR DECLARATIONS}
\subsection*{Conflict of Interest}
The authors have no conflicts to disclose.
\section*{Data Availability Statement}
The data that support the findings of this study are available from the corresponding author upon reasonable request.
\bibliography{References}
\end{document}